\documentclass[conference]{IEEEtran}
\IEEEoverridecommandlockouts
\usepackage{cite}
\usepackage{amsmath,amssymb,amsfonts}
\usepackage{algorithmic}
\usepackage{graphicx}
\usepackage{textcomp}
\usepackage{xcolor}
\usepackage{multirow}
\usepackage{placeins}
\def\BibTeX{{\rm B\kern-.05em{\sc i\kern-.025em b}\kern-.08em
    T\kern-.1667em\lower.7ex\hbox{E}\kern-.125emX}}

\begin{document}

\title{On Perception of Prevalence of Cheating and Usage of Generative AI}

\author{Roman Denkin\\
\textit{}\\
\textit{Department of Information Technology} \\
\textit{Uppsala University}\\
Uppsala, Sweden \\
roman.denkin@it.uu.se
}
\maxdeadcycles=1000
\extrafloats{1000}
\maketitle

\newpage
\begin{abstract}
This report investigates the perceptions of teaching staff on the prevalence of student cheating and the impact of Generative AI on academic integrity. Data was collected via an anonymous survey of teachers at the Department of Information Technology at Uppsala University and analyzed alongside institutional statistics on cheating investigations from 2004 to 2023. The results indicate that while teachers generally do not view cheating as highly prevalent, there is a strong belief that its incidence is increasing, potentially due to the accessibility of Generative AI. Most teachers do not equate AI usage with cheating but acknowledge its widespread use among students. Furthermore, teachers' perceptions align with objective data on cheating trends, highlighting their awareness of the evolving landscape of academic dishonesty. 
\end{abstract}

\section{Introduction}
The topic of student cheating has always been a contentious issue in educational environments. Despite the implementation of various punitive measures designed to deter students from cheating, the problem persists across all levels of education. Furthermore, technological advancements have introduced new methods of cheating that are both more difficult to detect and more effective. The latest controversial development in this area is Generative AI, which can be used to generate text for various purposes, including solving problems and writing scientific texts automatically. While the use of Generative AI is not universally defined as cheating, it represents an emerging issue that warrants investigation.

In this research, we aim to explore the perceptions of current teaching staff regarding the prevalence of cheating in education, including how these perceptions have changed over the years. Additionally, we will examine teachers' opinions on the use of Generative AI in education, including whether it should be considered cheating and their estimates of how many students currently use it. Understanding these subjective views is important because objective cheating statistics can be influenced by how skillfully students cheat, how effectively teachers or examiners can detect such activities, and the willingness of educational personnel to officially address these incidents.

\section{Literature Review}
The rapid advancement of technology and the increasing use of Generative AI in academia have brought new challenges to maintaining academic integrity. This review explores the current research on the detectability of AI-generated texts, the opportunities and challenges posed by AI in education, and the frequency and perception of cheating among students and faculty.

Fleckenstein et al. (2024) conducted a study on the ability of teachers to detect AI-generated texts among student essays. Their findings reveal that both novice and experienced teachers struggle to identify texts generated by ChatGPT, highlighting the difficulties in maintaining academic integrity with the rise of Generative AI \cite{FLECKENSTEIN2024100209}.

Cotton, Cotton, and Shipway (2024) discuss the dual nature of AI tools like ChatGPT in higher education. While these tools offer benefits such as increased engagement and accessibility, they also raise significant concerns about academic honesty and plagiarism. The paper suggests strategies for universities to ensure ethical use of AI, including policy development, training, and enhanced detection methods \cite{cotton2024chatting}.

Parnther (2020) provides a comprehensive review of academic misconduct in higher education. The study emphasizes the importance of academic integrity education and the need for clear guidelines and equitable resolutions. Parnther highlights the changing landscape of academic integrity education over time, including policy revisions and the role of various stakeholders \cite{parnther2020academic}.

In this paper, we will examine the problem from the perspective of teachers' perceptions of academic dishonesty and the challenges posed by Generative AI.

\section{Methodology}

\subsection{Data Sources}

To investigate the perception of cheating prevalence among teaching staff and the role of Generative AI in academic dishonesty, this research utilizes two primary sources of data:

\begin{enumerate}
    \item \textbf{Anonymous Teachers Survey}: Conducted within the Department of Information Technology at Uppsala University, Sweden.
    \item \textbf{Student Cheating Investigation Statistics}: Obtained from the Legal Administration of Uppsala University, detailing the number of cheating investigations conducted annually from 2004 to 2023.
\end{enumerate}

\subsection{Survey Instrument}

The survey was designed to capture the subjective views of teaching staff regarding the prevalence of student cheating and the impact of Generative AI on academic integrity. The survey included the following questions:

\begin{itemize}
    \item \textbf{Teaching Experience}:
    \begin{itemize}
        \item "What is your teaching experience in years?"
    \end{itemize}
    \item \textbf{Prevalence and Trends in Cheating} (using a Likert scale from 1 to 5):
    \begin{itemize}
        \item "Please estimate, how prevalent is students’ cheating nowadays?" (Scale: 1 - "It happens very rarely" to 5 - "It happens very often")
        \item "Do you think that students cheat more or less nowadays compared to the start of your teaching career?" (Scale: 1 - "It happens significantly less now" to 5 - "It happens significantly more often now")
    \end{itemize}
    \item \textbf{Generative AI and Cheating} (using a Likert scale from 1 to 5):
    \begin{itemize}
        \item "Do you think that usage of Generative AI to perform writing tasks is cheating?" (Scale: 1 - "It is not cheating at all" to 5 - "It is absolutely cheating")
        \item "How many of the students use Generative AI to complete their tasks, how would you estimate?" (Scale: 1 - "It happens very rarely" to 5 - "It happens very often")
    \end{itemize}
    \item \textbf{Open-Ended Question}:
    \begin{itemize}
        \item "If you would like to leave a special opinion on the prevalence of cheating in modern education and the importance of AI for this issue, please write your thoughts below."
    \end{itemize}
\end{itemize}

\begin{figure*}
\centerline{\includegraphics[width=0.7\textwidth]{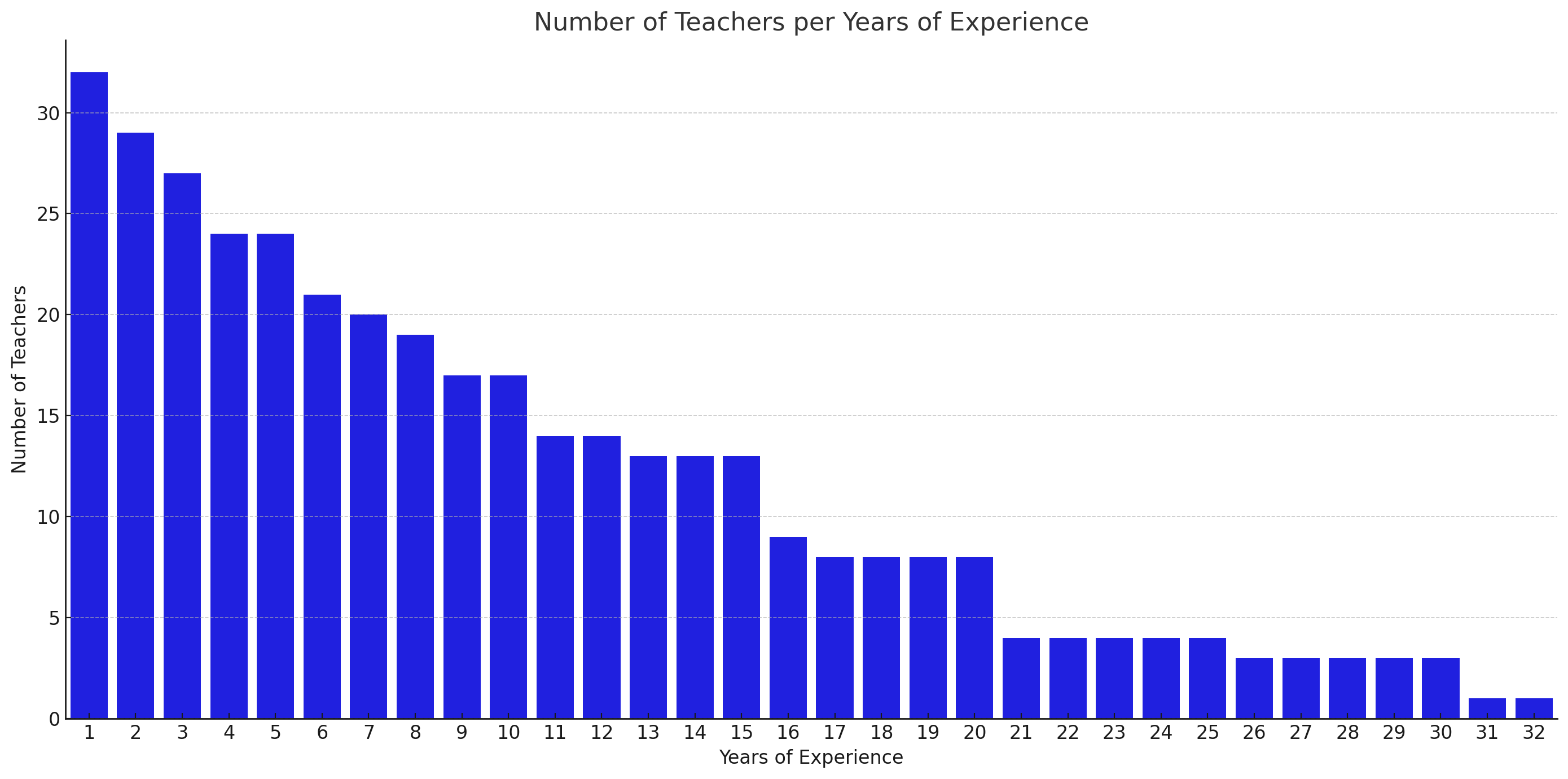}}
\caption{Distribution of teachers number per years of experience (cumulative)}
\label{fig:teachers}
\end{figure*}

The survey received 32 responses from teachers with teaching experience ranging from 1 to 32 years as depicted on Fig. \ref{fig:teachers}.

\subsection{Data Analysis}

The analysis will focus on the following aspects:

\begin{enumerate}
    \item \textbf{Descriptive Statistics}:
    \begin{itemize}
        \item Summarize the responses to each question to provide an overview of teachers' perceptions regarding cheating prevalence and the use of Generative AI in academic tasks.
    \end{itemize}
    \item \textbf{Trend Analysis}:
    \begin{itemize}
        \item Compare the perceived trend in cheating prevalence (from the survey responses) to the objective statistics on the number of cheating investigations conducted annually by the university.
    \end{itemize}
    \item \textbf{Experience-Based Comparison}:
    \begin{itemize}
        \item Analyze how the perception of cheating prevalence and the use of Generative AI varies between teachers with different levels of experience. Teachers will be grouped into two categories: those with less than 5 years of teaching experience and those with more than 5 years of teaching experience.
    \end{itemize}
\end{enumerate}

By combining subjective survey responses with objective institutional data, this methodology aims to provide a comprehensive understanding of the perception of current landscape of academic dishonesty and the emerging role of Generative AI in education. The findings will shed light on whether the perceived increase in cheating aligns with actual reported cases and how educators view the impact of advanced technology on academic integrity.

\section{Results}

\subsection{General Perception}

\begin{figure*}
\centerline{\includegraphics[width=1.0\textwidth]{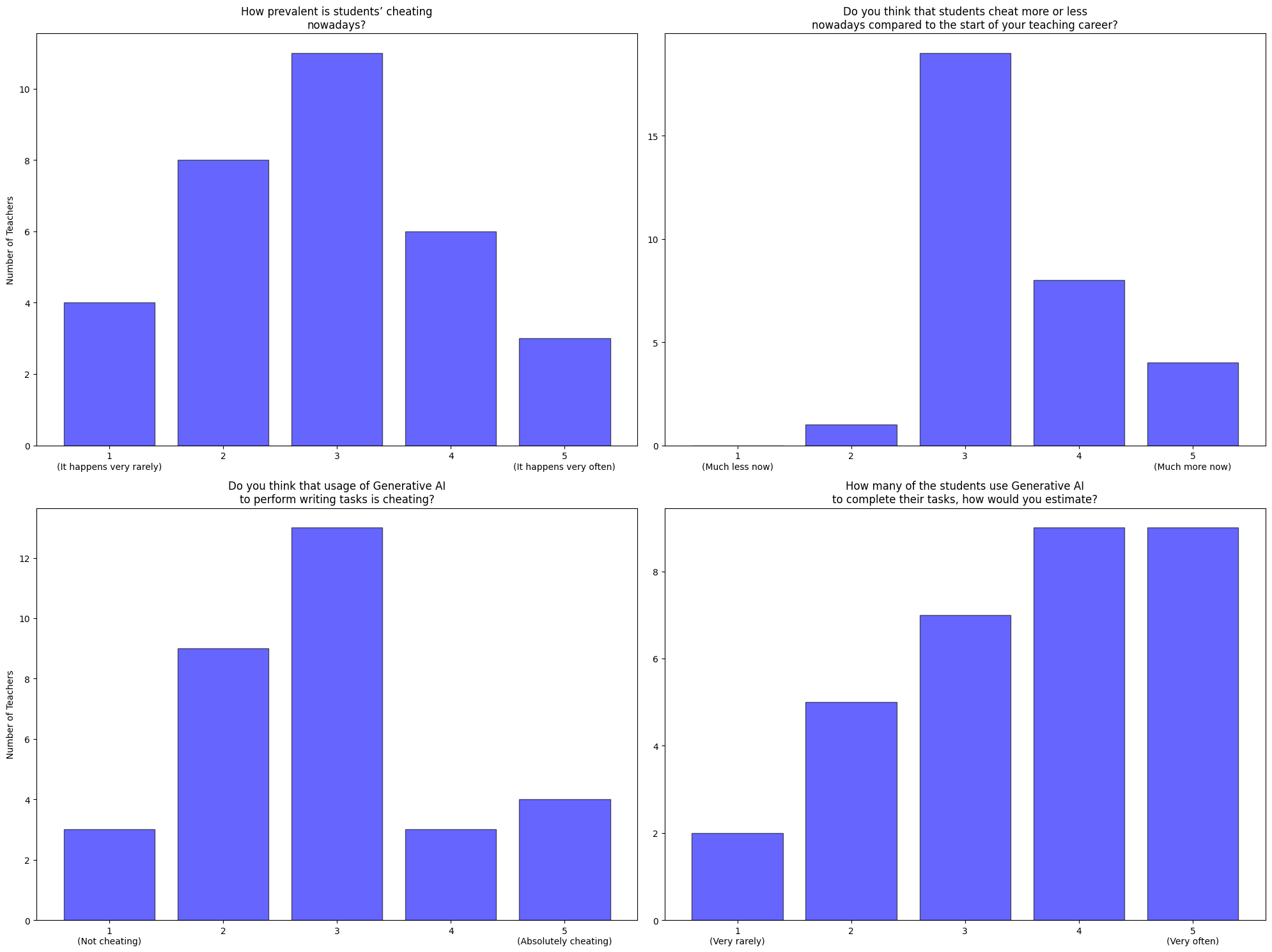}}
\caption{Responses of full group of teachers (1 to 32 years of experience)}
\label{fig:full}
\end{figure*}

The general distribution of answers is summarized in Figure~\ref{fig:full}. The survey results indicate that while teachers tend to believe that student cheating is not very common, with the majority rating its prevalence around the midpoint of the scale, there is a significant concern regarding the trend. Most teachers perceive an increase in cheating compared to the start of their teaching careers. Additionally, opinions on whether the use of Generative AI constitutes cheating tend to lean slightly below neutral, suggesting that many do not view it as outright cheating. However, there is a strong consensus that the use of AI by students to complete tasks is quite prevalent nowadays.

\subsection{Cheating Prevalence Trends}

\begin{figure*}
\centerline{\includegraphics[width=0.7\textwidth]{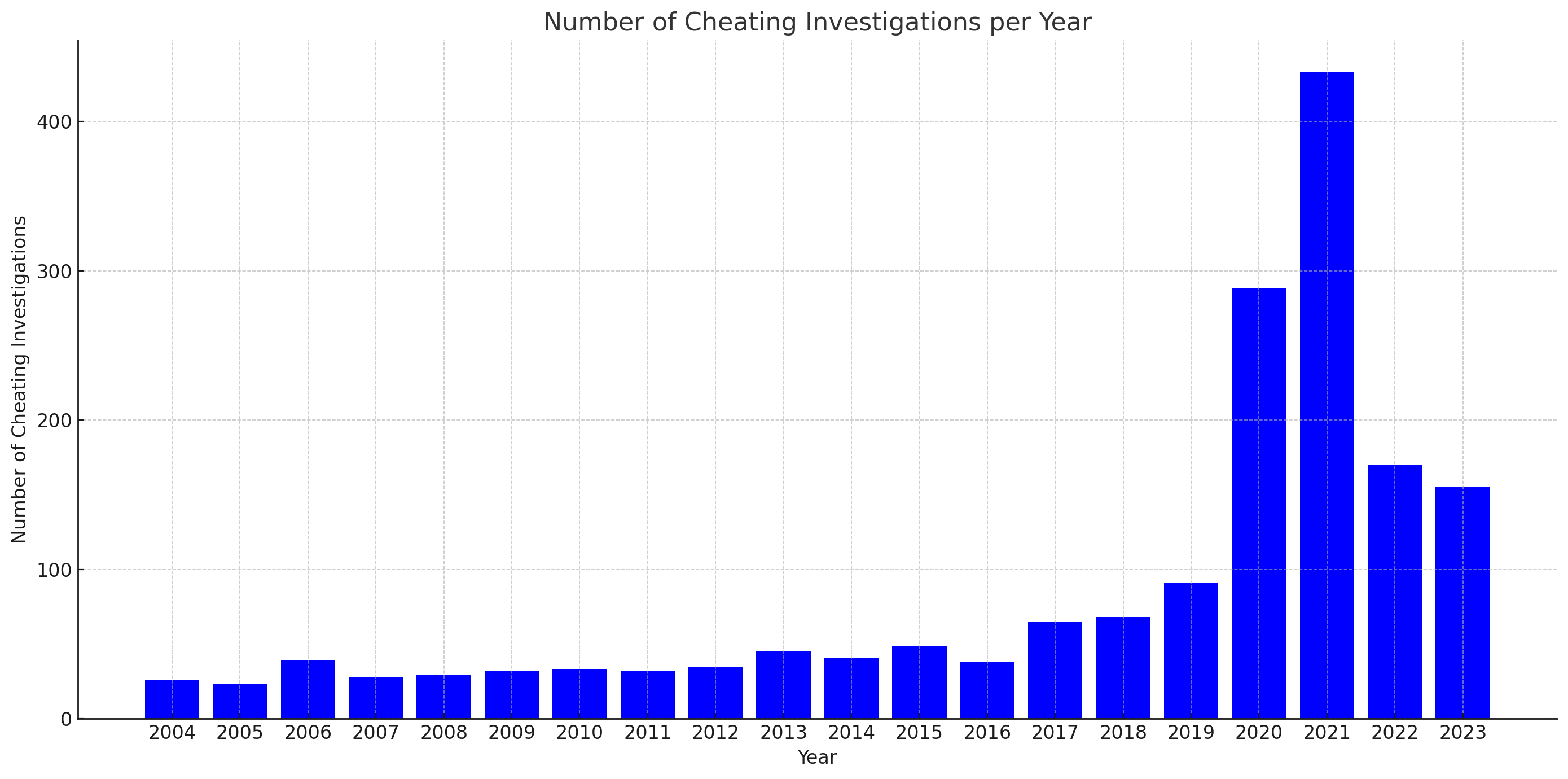}}
\caption{Number of Cheating Investigations per Year}
\label{fig:investigations}
\end{figure*}

\begin{figure*}
\centerline{\includegraphics[width=0.7\textwidth]{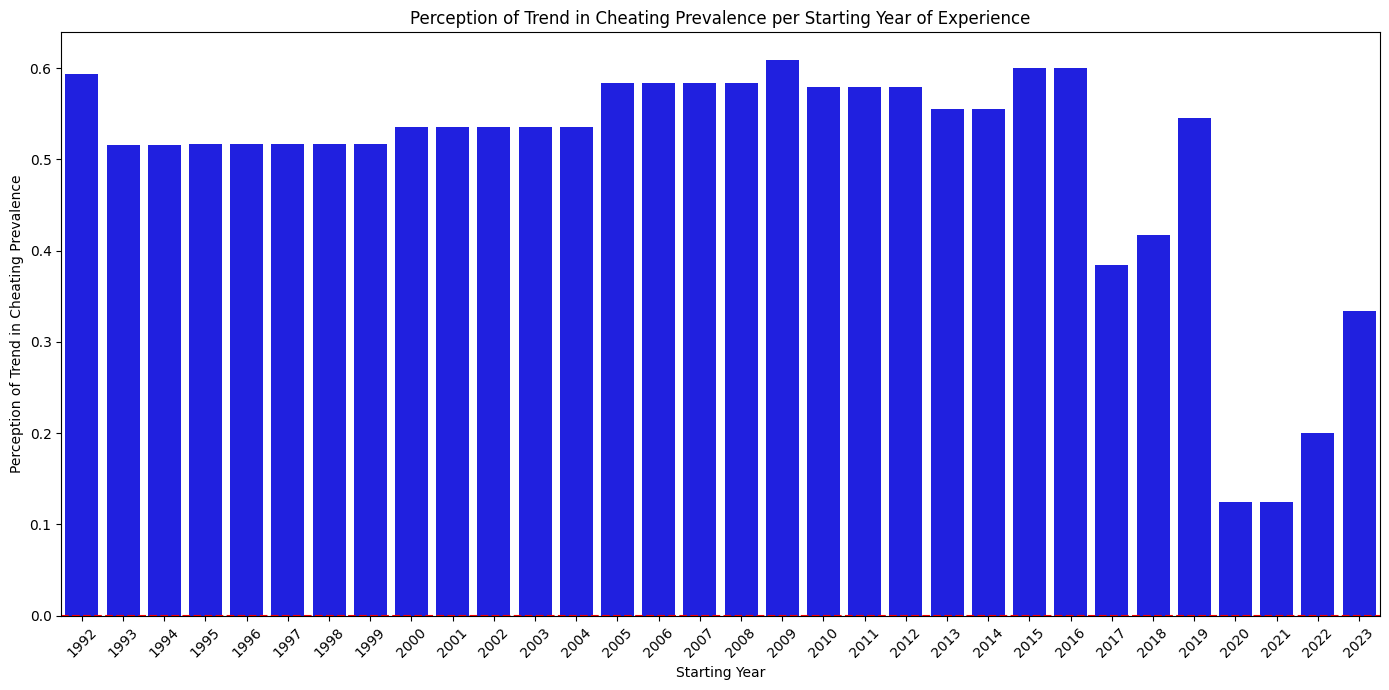}}
\caption{Perception of Trend in Cheating Prevalence per Starting Year of Experience}
\label{fig:perception}
\end{figure*}

Figure~\ref{fig:investigations} illustrates the number of cheating investigations conducted annually at Uppsala University from 2004 to 2023. The data reveals a noticeable increase in the number of investigations over the years, with a significant peak during the COVID-19 pandemic years (2020-2021).

Figure~\ref{fig:perception} shows the average perception scores of cheating prevalence trends, remapped to a scale from -3 to +3, with positive values indicating a perceived increase in cheating. The perception scores are averaged for teachers who started their careers in a specific year or later. The results correlate with the objective statistics, indicating that teachers correctly perceive trends in cheating prevalence over time.

Additionally, a significant peak in cheating cases during the COVID-19 pandemic years is also evident in the perception scores. Teachers who started their careers during these years perceive a significantly lower increase in cheating frequency, indicating that their experience during the pandemic has influenced their views on the trend of cheating prevalence.

\subsection{Difference in Experience Groups' Perception}

\begin{figure*}[ht!]
\centerline{\includegraphics[width=1.0\textwidth]{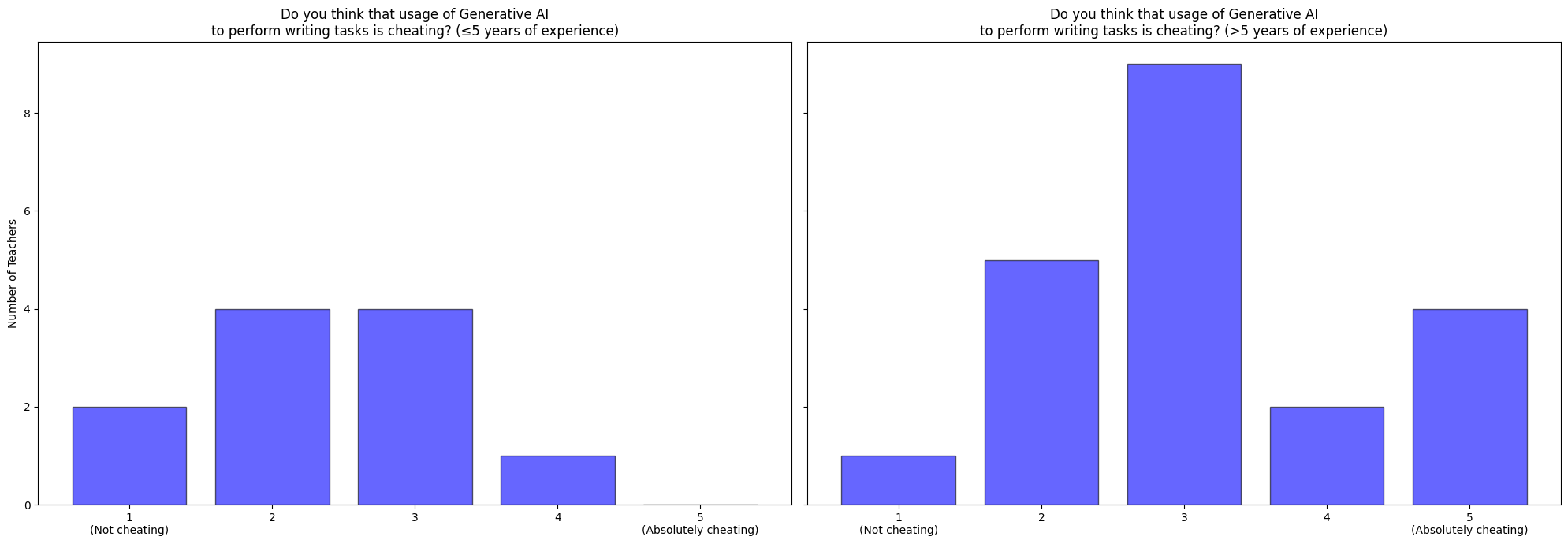}}
\caption{Perception of AI usage as cheating by experience group}
\label{fig:ai_cheating_age}
\end{figure*}

\begin{figure*}[ht!]
\centerline{\includegraphics[width=1.0\textwidth]{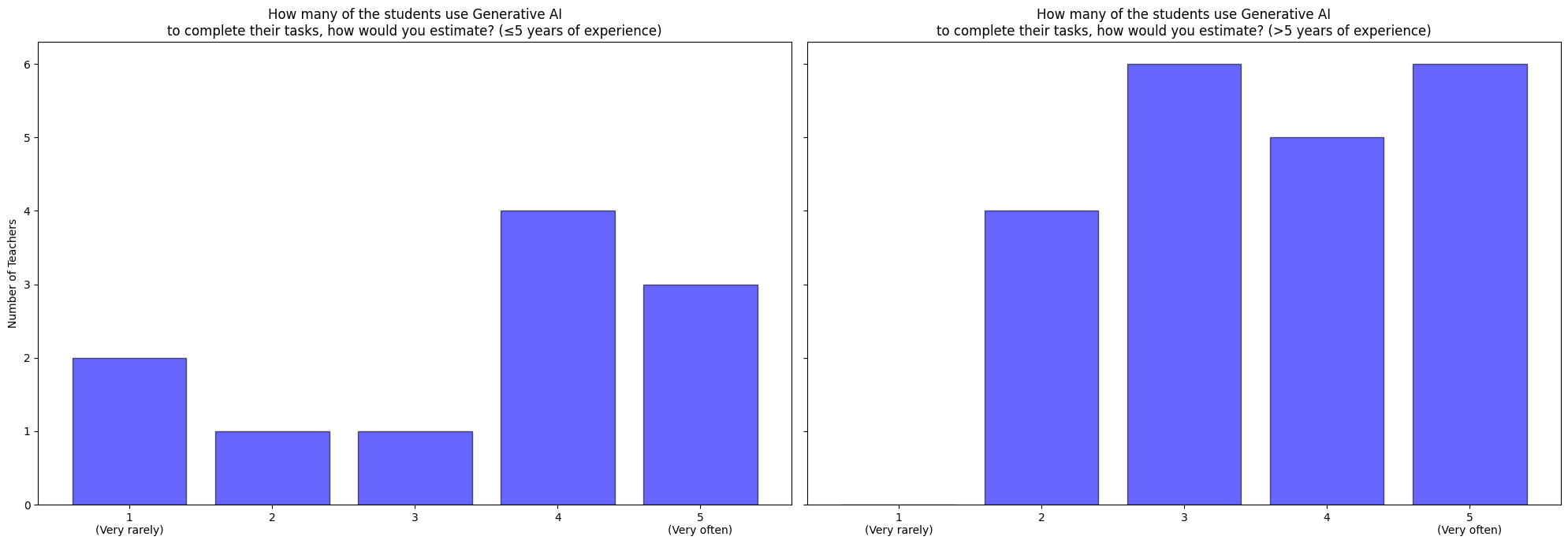}}
\caption{Perception of AI usage prevalence by experience group}
\label{fig:ai_usage_age}
\end{figure*}

In this section, we compare the survey responses between two groups of teachers: those with less than 5 years of teaching experience and those with more than 5 years of teaching experience. The detailed data can be seen in the Appendix (Figures~\ref{fig:cheating_age}, \ref{fig:trend_age}, \ref{fig:ai_cheating_age}, and \ref{fig:ai_usage_age}).

On the current prevalence of cheating, both groups of teachers generally believe that cheating is not very prevalent (Figure~\ref{fig:cheating_age}). However, the more experienced group exhibits a wider distribution of answers, indicating that some teachers in this group do believe that cheating is very common.

Regarding the trend in cheating prevalence, most teachers believe that the prevalence of cheating is increasing, with no significant differences between the experience groups (Figure~\ref{fig:trend_age}).

Similarly, in terms of the use of Generative AI, both groups tend to believe that its usage does not necessarily constitute cheating (Figure~\ref{fig:ai_cheating_age}). However, the more experienced group has instances of stronger negative opinions on this matter.

Finally, both groups agree that Generative AI is widely used by students to perform their tasks, with the more experienced group holding stronger opinions about the extent of its use (Figure~\ref{fig:ai_usage_age}).

\section{Conclusion}

The results of our survey and analysis reveal several key insights into teachers' perceptions of cheating and the impact of Generative AI in education:

\begin{enumerate}
    \item While teachers generally do not believe that cheating is very common, there is a strong opinion that the prevalence of cheating is increasing. This may be attributed to the introduction of Generative AI to the general public, making powerful cheating tools more accessible.
    \item Teachers do not necessarily equate the usage of Generative AI with cheating, but most agree that a significant portion of students do use it.
    \item Teachers' perceptions of cheating align with objective data on the frequency of cheating incidents, indicating a reflection of actual trends.
\end{enumerate}

Furthermore, teachers have strong opinions on the issue of the appearance of Generative AI and its impact on teaching. Selected quotes covering the spectrum of these opinions are included in the Appendix.

\section{Limitations and Future Research}

Due to the limitations of this research project, the teachers' survey was made very concise and was sent to a subject-limited group of teachers from the Department of Information Technology at Uppsala University. Factors such as gender, age, and other possible demographic biases were not measured or balanced, as this data was not available in the survey.

Future research could involve conducting a broader survey that includes a more diverse audience, covering different departments and institutions. Additionally, obtaining demographic data would help to balance potential biases and provide a more comprehensive understanding of teachers' perceptions across various groups.

\clearpage
\onecolumn
\FloatBarrier
\section*{Appendix I: Experience Groups Responses}

\begin{figure*}[ht!]
\centerline{\includegraphics[width=1.0\textwidth]{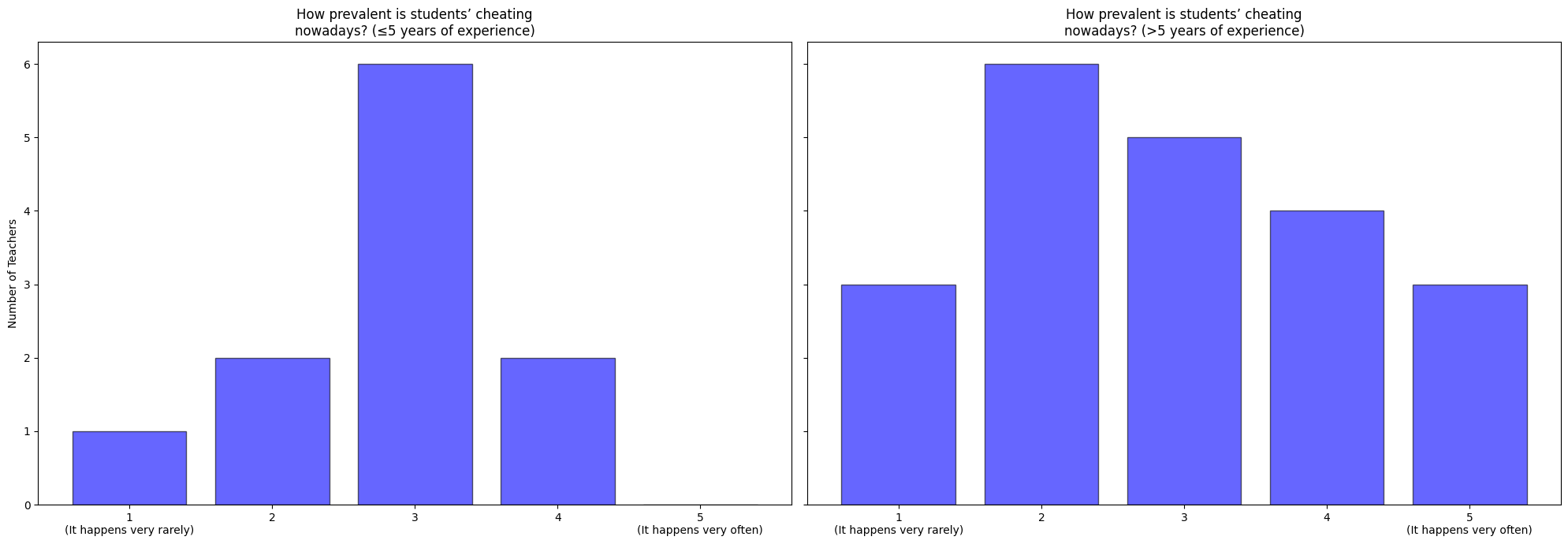}}
\caption{Perception of current cheating prevalence by experience group}
\label{fig:cheating_age}
\end{figure*}

\begin{figure*}[ht!]
\centerline{\includegraphics[width=1.0\textwidth]{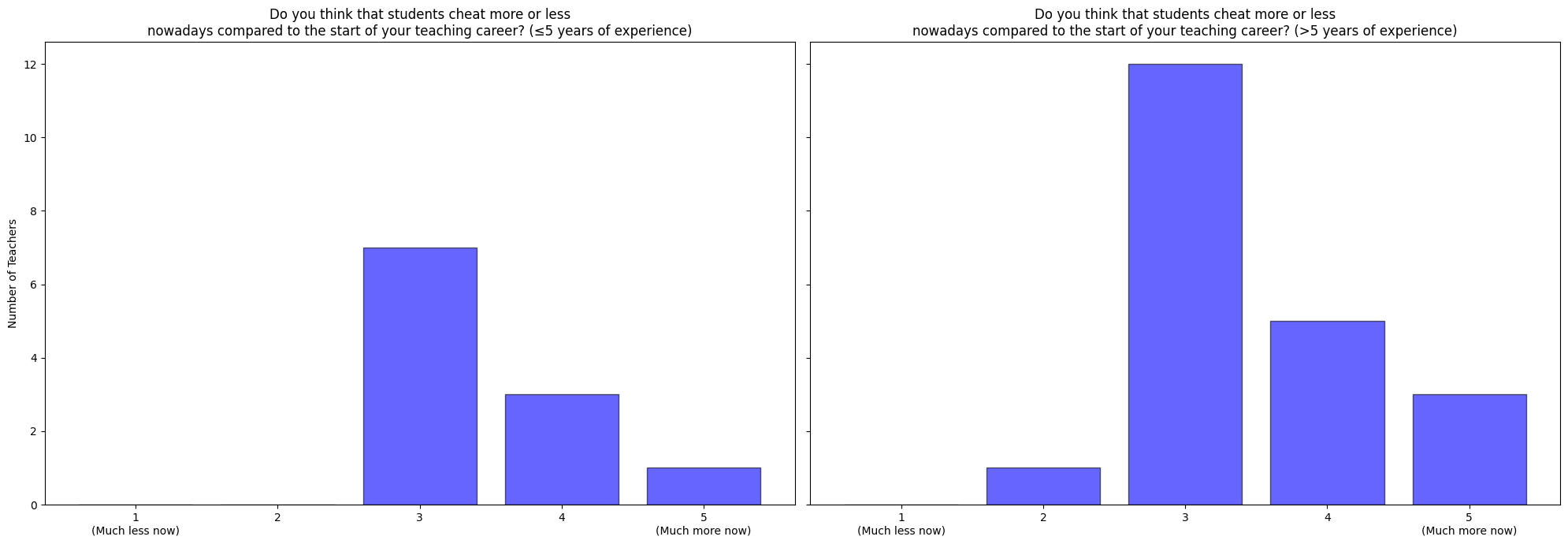}}
\caption{Perception of trend in cheating prevalence by experience group}
\label{fig:trend_age}
\end{figure*}

\clearpage
\FloatBarrier
\section*{Appendix II: Selected Quotes}

\textbf{Opinions considering AI as a problem:}

\begin{quote}
    "I think it is important for teachers to talk about it as a tool and how it can be used. There are enough limitations with it that it is possible to create assignments where it can't all be completed by ChatGPT, for example. Or create assignments that assume the involvement of AI. AI can help people with language barriers and without proofreaders to support writing. However, I think it can also be a problem for people with limited English to then not be able to catch a lot of the AI errors and then turn in poor or unrelated work, but just written more clearly. Also in writing, there needs to be references, so I think it raises the importance of checking references to ensure the student has put in the work. To say that it shouldn't be used at all puts a lot of people in a moral grey area. Publications are also allowing some use of ChatGPT for abstracts and proofreading, so it's another reason why it seems reasonable in writing."
\end{quote}

\vspace{0.5cm}

\begin{quote}
    "Using generative AI for writing is not a problem in the fields I teach (programming and automatic control). Most students hand in handwritten solutions of math derivations. And in software design, the models are not good enough to really use in a way that would be cheating. But in smaller programming exercises, where previous years' hand-ins are on GitHub, GitHub Copilot can solve the solutions. This is a problem, as a computer-generated hand-in and a handwritten hand-in will likely be identical. We need to move the examination from just producing an answer into more higher-level tasks, such as explaining the solution and discussing the implications of it. This means that we are forced to examine something different than we initially wanted. I see this as a problem. Minor at the moment, but still a problem. And growing."
\end{quote}

\vspace{0.5cm}

\textbf{Opinions considering AI as not a problem:}

\begin{quote}
    "I believe that using AI as a TOOL is not considered cheating."
\end{quote}

\vspace{0.5cm}

\begin{quote}
    "The use of AI can be a grey area, not always easy to rule out legit use."
\end{quote}

\vspace{0.5cm}

\textbf{Opinions suggesting a need to change teaching methods:}

\begin{quote}
    "I think there are valid parallels to be made with the introduction of calculators in the 70s (which was a huge controversy at the time): if we keep evaluating students on the same skills as before, in the same way as before, then they just gained a powerful tool that lets them get good grades without engaging their brain or attaining the intended learning outcomes. Some change in our goals and practices needs to happen; what the change might be is too early to say."
\end{quote}

\vspace{0.5cm}

\begin{quote}
    "I think we need to adapt the tasks."
\end{quote}

\end{document}